\documentstyle[aps,preprint,prl,citesort,epsf]{revtex}

\newcommand{\eqref}[1]{Eq.~(\protect\ref{#1})}
\newcommand{\figref}[1]{Fig.~\protect\ref{#1}}

\begin{document}

\draft

% title-and-abstract.tex

\title{Karhunen-Lo\`eve Decomposition of Extensive Chaos}

\author{
Scott~M.\ Zoldi\cite{CNCS-address}\cite{zoldi-email} and
Henry~S.\ Greenside\cite{CNCS-address}
}

\address{
Department of Physics\\
Duke University, Durham, NC 27708-0305
}

\date{October 1, 1996}

\maketitle

% Max of 600 characters for PRL abstract

\begin{abstract}
We show that the number of KLD (Karhunen-Lo\`eve
decomposition) modes $D_{\rm KLD}(f)$ needed to capture
a fraction~$f$ of the total variance of an extensively
chaotic state scales extensively with subsystem
volume~$V$. This allows a correlation length~$\xi_{\rm
KLD}(f)$ to be defined that is easily calculated from
spatially localized data. We show that~$\xi_{\rm
KLD}(f)$ has a parametric dependence similar to that
of the dimension correlation length and demonstrate
that this length can be used to characterize
high-dimensional inhomogeneous spatiotemporal chaos.
\end{abstract}


\narrowtext

Spatiotemporal states that are nonperiodic both in
space and time abound in nature and are often important
technologically, e.g, for lasers, fibrillating hearts,
and convective transport of heat
\cite{Cross93}. Experiments and simulations 
of these dynamics raise questions that are presently
poorly understood: Are there different kinds of
spatiotemporal nonperiodic states? What kinds of
bifurcations lead to such states? How do
inhomogeneities and boundary conditions affect the
dynamics? And how does the transport of energy and
matter depend on the details of the spatiotemporal
disorder?  An essential first step towards answering
these questions is to develop methods to quantify
spatiotemporal dynamic states, so that one state can be
distinguished from another and so that theory can be
compared with experiment and with simulation. Data
analysis and theoretical progress are presently limited
by a scarcity of concepts and of computational tools
for analyzing spatiotemporal disorder.

In the absence of a fundamental theory of sustained
nonequilibrium systems that could indicate appropriate
quantities to measure, researchers have used primarily
two approaches for quantifying spatiotemporal disorder:
two-point correlation functions and dynamical
invariants such as Lyapunov exponents and fractal
dimensions \cite{Ott93}. Temporal correlation functions
have been effective for distinguishing periodic and
quasiperiodic dynamical states from each other and from
chaotic ones \cite{Ahlers78ptps} while spatial
correlation functions have played an important role in
discovering and demonstrating the absence of long range
spatial order
\cite{decay-of-spatial-correlations}. Both correlation
functions have been less useful for distinguishing one
chaotic state from another or for comparing
experimental with computational chaotic data. Dynamical
invariants have been somewhat useful for ordering the
chaotic states of low-dimensional dynamical systems but
severe difficulties remain in calculating these
quantities for higher-dimensional systems because of
the demanding computational effort, the slow and often
ambiguous convergence of time-series-based algorithms,
and the need for large amounts of noise-free
data~\cite{Abarbanel93}.

In this Letter, we propose and analyze a new measure of
spatiotemporal disorder, a correlation length~$\xi_{\rm
KLD}$ defined below, that seems promising especially
for the analysis of large, high-dimensional,
nontransient, driven-dissipative systems. This quantity
has some of the flavor of correlation functions and
also of dynamical invariants and is straightforward to
compute with moderate amounts of spatiotemporal data
(unlike global dynamical invariants like the fractal
dimension) since, as we show below, it is a {\em local}
quantity that can be calculated from spatiotemporal
data associated with a finite region of space. This
last feature suggests that~$\xi_{\rm KLD}$ will be
useful for studying spatially inhomogeneous dynamics
arising from slowly changing external parameters, from
broken symmetries \cite{Sasa96}, or from the influence
of boundaries \cite{Gluckman95}. As is the case for
many measures of spatiotemporal chaos, the length
$\xi_{\rm KLD}$ can not be evaluated analytically and
so we investigate its properties numerically for two
idealized mathematical models---the one-dimensional
Kuramoto-Sivashinsky (KS) equation
\cite{Manneville88,Cross93} and the two-dimensional
Miller-Huse model \cite{Miller93}---whose
spatiotemporal chaotic solutions have been thoroughly
studied and for which inhomogeneities can be introduced
in a controlled manner. In later work, applications to
experimental data will be reported.

Our motivation for defining and studying the
correlation length~$\xi_{\rm KLD}$ comes from three
different ideas. First is the idea of extensive chaos,
that a sufficiently big {\em homogeneous}
spatiotemporal chaotic system has the property that its
fractal dimension~$D$ is extensive, growing in
proportion to the system volume~$V$
\cite{Ruelle82,Manneville85,Cross93,Egolf95prl}. This
extensive scaling suggests that bounded {\em intensive}
quantities, such as a dimension density~$\delta =
\lim_{V\to \infty} D/V$ or the equivalent
dimension correlation length~$\xi_\delta =
\delta^{-1/d}$ \cite{Cross93,Egolf95prl,OHern96}
(where~$d$ is the number of asymptotically large
spatial dimensions, e.g., $d=2$ for a
large-aspect-ratio convection experiment) are more
appropriate for characterizing large nonequilibrium
systems. It was shown recently that the
length~$\xi_\delta$ varies independently of the
two-point and mutual-information correlation lengths so
that the average spatial disorder does not determine
the fractal dimension~$D$ \cite{OHern96}. This suggests
that certain measures of spatiotemporal dynamics should
be sensitive to structure in phase space, not just to
instantaneous \cite{Gunaratne95} or time-averaged
measures of spatial disorder in configuration
space. The quantity~$\xi_{\rm KLD}$ turns out to have
this property.

The second idea is to extend the concept of local
thermodynamic equilibrium \cite[page 13]{Landau80} to
slowly-varying inhomogeneous driven dissipative
systems, with the implication that intensive dynamical
quantities can be defined locally and will be slowly
varying in space. As an illustration, assume that a
large sustained nonequilibrium systems has a
parameter~$p(x)$ that varies slowly with position~$x$
and consider a subsystem of size~$L$ centered at
position~$x_0$. Then over a certain time scale that
decreases with decreasing subsystem size~$L$ (not
necessarily diffusively), this subsystem will be
approximately nontransient, even if the system
containing the subsystem is not. Further, the values of
intensive parameters associated with this approximately
nontransient subsystem should correspond closely in
value to those of an infinite homogeneous nontransient
system with the parameter value~$p(x_0)$.
Unfortunately, there are no reliable algorithms that
can estimate intensive quantities such as the dimension
density~$\delta$ from information localized to some
region of space (it is not even known whether such
algorithms exist in principle). To date, calculations
of intensive quantities such as the Lyapunov dimension
density~$\delta$ have relied on the expensive
calculation of global extensive quantities followed by
taking the limit of some intensive ratio
\cite{OHern96}. This is impractical for the analysis
of experimental data or for the evaluation of local
measures. As we show below, the measure~$\xi_{\rm KLD}$
is able to indicate some of the behavior of the fractal
dimension density using just local information.

The third idea is to make use of the Karhunen-Lo\`eve
decomposition (KLD), which has been used by researchers
over many years and in many disciplines to analyze
spatiotemporal data \cite{kld-refs}, although not in
the context of extensive chaos or of inhomogeneous
systems. The KLD is a statistical method for
compressing spatiotemporal data by finding the largest
linear subspace that contains substantial statistical
variations of the data.  To illustrate the idea in the
discrete case and also to introduce some notation, we
consider a one-dimensional zero-mean field~$u(t,x)$ on
a spatial interval~$[0,L]$ whose values are measured on
a finite space-time mesh of $T$~uniformly sampled time
points~$t_i = i
\triangle{t}$ and of $S$~uniformly sampled spatial
points~$x_j = j \triangle{x}$.
%(Generally~$T \gg S$ since it is often easier to
% measure long time series at a few points in space
% in a given experiment.)
Then a $T\times S$ rectangular data matrix~$A_{ij} =
u(t_i,x_j)$ can be defined from which a $S \times S$
symmetric positive semidefinite scatter matrix ${\bf M}
= {\bf A}^T {\bf A}$ can be calculated, where~${\bf
A}^T$ denotes the matrix transpose of~$\bf A$. Using
standard eigenvalue methods with complexity~$O(S^3)$
\cite{LAPACK95}, the scatter matrix can be diagonalized
to obtain its nonnegative eigenvalues~$\sigma_i^2$
which can be further ordered in decreasing size~$\sigma_1^2
\ge \sigma_2^2 \cdots \ge \sigma_S^2 \ge 0$.

Since the ordered eigenvalues~$\sigma_i^2$ often
decrease rapidly in magnitude with increasing
index~$i$, researchers \cite{kld-dimension-refs} have
introduced a positive integer~$D_{\rm KLD}(f)$:
\begin{equation}
  D_{\rm KLD}(f) = \max \left\{
    p : \, \sum_{i=1}^p \sigma_i^2
        \bigg/
           \sum_{i=1}^S \sigma_i^2 \, \le f
  \right\}
  , \label{eq:D-KLD-defn}
\end{equation}
which represents the number of KLD~modes needed to
capture some specified fraction~$f \le 1$ of the total
variance~$\sum_{i=1}^S \sigma_i^2$ of the
data. Researchers have suggested using~$D_{\rm KLD}(f)$
like a fractal dimension~$D$ to measure the complexity
of spatiotemporal data \cite{kld-dimension-refs}
although care is needed when interpreting~$D_{\rm
KLD}(f)$. The~$T$ $S$-dimensional vectors defined by
the rows of the data matrix~$\bf A$ constitute an
embedding of the dynamics into a~$S$-dimensional phase
space. The quantity $D_{\rm KLD}(f)$ indicates the
dimension of a linear subspace that includes most of
the statistical variation of this embedding and is
generally quite different from the attractor's fractal
dimension~$D$, e.g., a limit cycle with fractal
dimension~$D=1$ in a~$N$-dimensional phase space could
have a value of~$D_{\rm KLD}(f)$ between~1 and~$N$
depending on how the limit cycle is folded in different
orthogonal directions.

Although~$D_{\rm KLD}(f)$ need have no particular
relation to the data's fractal dimension~$D$, we argue
below that for {\em extensive} chaos, the rate of
increase of~$D_{\rm KLD}(f)$ with volume~$V$ will
generally be similar to the rate of increase of fractal
dimension~$D$ with~$V$ since the extensivity of the
dimension arises from the appearance of new orthogonal
directions, i.e., with increasing~$V$, a larger linear
space is needed to contain the increasing variance of
of the attractor. If true, we can then use the more
readily calculated quantity~$D_{\rm KLD}(f)$ to
estimate intensive correlation lengths like the
length~$\xi_\delta$ discussed above, and to study the
dependence of such lengths on system parameters.

We now present some results that justify and illustrate
these observations.  When evaluated for spatiotemporal
data of a large, approximately homogeneous, sustained
nonequilibrium system of volume~$V$, the KLD
dimension~$D_{\rm KLD}(f)$ of~Eq.~(\ref{eq:D-KLD-defn})
grows extensively with~$V$ as shown in
\figref{fig:KLD-dim-extensive}. Here we have used
$T \times S$ data matrices~$\bf A$ (with~$10^4 \le T
\le 2 \times 10^4$ and $100 \le S \le 800$) derived
from the spatiotemporal field~$u(t,x)$ of the
one-dimensional Kuramoto-Sivashinsky equation
\begin{equation}
      \partial_t u
    + \partial_x^2 u  + \partial_x^4 u
    + u \, \partial_x u= 0, \qquad x \in [0, L]
  , \label{eq:KS-eq}
\end{equation}
with rigid boundary conditions~$u = \partial_x u = 0$
at~$x=0$ and at~$x=L$. \eqref{eq:KS-eq} was integrated
numerically with a semi-implicit finite-difference
method that was first- and second-order accurate in
time and space respectively. For~$L > 50$, most initial
conditions yield spatiotemporal chaotic states that
were previously shown to be extensively chaotic
\cite{Manneville85}. Fig~\ref{fig:KLD-dim-extensive}(a)
shows that~$D_{\rm KLD}(f)$ is extensive for~$L \ge
50$, growing in proportion to the system volume~$L$
with a slope that depends on the fraction~$f$. The
dashed line indicates the extensive scaling of the
Lyapunov dimension~$D$ \cite{Manneville85} which
corresponds to a fraction
$f=0.81$. Fig~\ref{fig:KLD-dim-extensive}(b) shows
further that~$D_{\rm KLD}(f)$ is extensive for open
subsystems centered on the middle of a system of
size~$L=400$. The slope of a line in
Fig~\ref{fig:KLD-dim-extensive}(b) for a given
fraction~$f$ is the {\em same} as the slope of the line
of corresponding~$f$ in
Fig~\ref{fig:KLD-dim-extensive}(a). This implies the
important point that the intensive density $\lim_{V \to
\infty} D_{\rm KLD}(f)/V$ can be estimated from
information localized to a certain region of space.

The extensivity of the KLD~dimension, for both the
entire system and for subsystems, suggests introducing
an intensive~KLD correlation length~$\xi_{\rm KLD}(f)$:
\begin{equation}
  \xi_{\rm KLD}(f) = \left(
   \lim_{V \to \infty} \frac{D_{\rm KLD}(f)}{V}
  \right)^{-1/d}
  , \label{eq:xi-KLD-defn}
\end{equation}
by analogy to the dimension correlation
length~$\xi_\delta$ (where again~$d$ is the spatial
dimensionality of the system).  Based on the data in
\figref{fig:KLD-dim-extensive} for the KS~equation,
\figref{fig:xi-KLD-vs-f} shows how the length
$\xi_{\rm KLD}(f)$ varies with the fraction~$f$. The
dependence is nonlinear, with the magnitude
of~$\xi_{\rm KLD}(f)$ changing by a factor of ten over
the range~$0.3 \le f \le 1$. Contrary to an earlier
claim by Ciliberto and Nicolaenko
\cite{kld-dimension-refs}, \figref{fig:xi-KLD-vs-f}
shows that the fractal dimension of a high-dimensional
system can not generally be estimated from a knowledge
of~$\xi_{\rm KLD}(f)$ since the fraction~$f$
corresponding to the dimension correlation length will
not be known in advance and because~$\xi_{\rm KLD}(f)$
can vary substantially with~$f$. However, the onset of
extensivity for~$D_{\rm KLD}(f)$ does accurately
predict the onset of extensivity for the Lyapunov
dimension~$D$ with increasing volume~$V$.

\figref{fig:xi-kld-parametric} shows
how the length~$\xi_{\rm KLD}(f)$ (for~$f=0.95$)
compares with the dimension correlation
length~$\xi_\delta$ (derived from the extensive
Lyapunov fractal dimension \cite{OHern96}) for a
nonequilibrium Ising-like phase-transition of a
mathematical model invented by Miller and Huse
\cite{Miller93}. The Miller-Huse (MH) model is a
2d~coupled-map-lattice for which a 1d~chaotic map of
odd symmetry is placed at each node of a periodic $L
\times L$ square lattice. Each site is
coupled diffusively to nearest neighbors
with a strength~$g$ that acts as the bifurcation
parameter for this system. Miller and Huse found that a
quantity analogous to a lattice magnetization~$M$
bifurcated from a zero to non-zero value at a critical
value~$g_c = 0.205$ at which point also a two-point
correlation length~$\xi_2$ diverged to infinity. O'Hern
et al \cite{OHern96} showed that the dimension
correlation length~$\xi_\delta$ did not diverge
near~$g_c$ but instead was a quantity of order one that
smoothly reached a local maximum at a value~$g=0.200$
distinctly less than the critical
value~$g_c$. \figref{fig:xi-kld-parametric} shows that
the length~$\xi_{\rm KLD}$ is somewhat larger
than~$\xi_\delta$ and has a qualitatively similar
dependence on~$g$ in that it increases to a local
maximum at the {\em same} $g$-value. Unlike the
dimension correlation length, $\xi_{\rm KLD}$ does not
seem to vary smoothly near its maximum but we lack
sufficient numerical resolution to determine
unambiguously whether there is a finite jump in value,
in analogy to the dependence of specific heat on
temperature for a second-order equilibrium phase
transition. The functional dependence of ~$\xi_{\rm
KLD}(f)$ on parameter~$g$ depends only weakly on the
fraction~$f$ which therefore is not an important
parameter.

Finally, in \figref{fig:xi-KLD-spatial-variation} we
demonstrate how the KLD correlation length~$\xi_{\rm
KLD}$ can be used to characterize inhomogeneous
spatiotemporal chaos, a result that opens up
interesting possibilities for the future analysis of
experimental data. For this figure, we introduced a
spatial inhomogeneity into a $300 \times 30$ periodic
MH~lattice by allowing the coupling constant~$g=g(x)$
to vary periodically in the~$x$ lattice direction as
shown in \figref{fig:xi-KLD-spatial-variation}(b). At
each of several $x$-coordinates $i$, we calculated the
KLD dimension \eqref{eq:D-KLD-defn} for subsystems
centered on~$i$ and of increasing width~$L$ with~$9 \le
L \le 15$. From these data, local extensive scaling was
identified from which a length~$\xi_{\rm KLD}(i)$ was
calculated from \eqref{eq:xi-KLD-defn}. In
\figref{fig:xi-KLD-spatial-variation}(a), the
lengths~$\xi_{\rm KLD}(i)$ are given by the solid
circles which can be compared with the dashed-curve
representing the corresponding value of~$\xi_{\rm KLD}$
that would be obtained
from~\figref{fig:xi-kld-parametric} for an infinite
homogeneous system with constant value~$g=g(i)$. The
agreement is good to about~4\% throughout, which is
sufficient to determine the spatial dependence of the
inhomogeneity.

%KLD yields both spectrum and modes but we focus mainly
%on the spectrum. We also ignore the time dependence of
%coefficients in the KLD which will be important to
%study later on.

In conclusion, the correlation length
\eqref{eq:xi-KLD-defn} obtained by studying the
extensive scaling of the Karhunen-Lo\`eve decomposition
with increasing subsystem volume provides an easily
calculated and novel way to characterize the
spatiotemporal disorder of an extensively chaotic
system, including the case of slowly varying spatial
inhomogeneities.  We believe that the ideas on which
our analysis is based---namely extensivity, local
stationarity, and the Karhunen-Lo\`eve
decomposition---will be important ingredients in the
future analysis of large nonequilibrium systems.

This work was supported by a DOE Computational Science
Fellowship, by NSF grants NSF-DMS-93-07893 and
NSF-CDA-92123483-04, and by DOE grant
DOE-DE-FG05-94ER25214.

% To get final references, uncomment \input{references}
% line and comment \biblio.. lines

\bibliographystyle{prsty}  % entries in order of citation

% Before submitting paper, comment out \bibliography
% line and uncomment out the \input{refences} line

%\bibliography{cardiology,chaos,na,hsg,physics,spatiotemporal,statistics,time-series}

% Figure captions here

\newpage
% figures.tex

\begin{figure}   % first figure
\caption{
  (a) KLD~dimension $D_{\rm KLD}(f)$
  \eqref{eq:D-KLD-defn} versus system size~$L$ for
  \eqref{eq:KS-eq} with rigid boundary conditions, for
  system sizes~$50 \le L \le 400$.  The percentage
  labels indicate the value of the fraction~$f$ for
  each curve.
% The integration algorithm used a
% space-time resolution of~$\Delta{t}=0.025$
% and~$\Delta{x}=0.25$.
% and initial data consisting of
% uniformly distributed random numbers in the interval
% $[-0.01,0.01]$.
  After integrating~$10^6$ time steps to allow
  transients to decay, $T=2\times 10^4$ time-samples
  with step~$\triangle{t}=2$ time units were
  used to construct the scatter matrix~$\bf M$. The
  dashed line shows the extensive Lyapunov dimension
  from Ref.~\protect\cite{Manneville85} which
  corresponds to a fraction~$f=0.81$.  (b) For
  the same numerical parameters and for a fixed system
  size of~$400$, a plot of $D_{\rm KLD}(f)$ versus
  subsystem size~$L$ of the~1d KS~equation.  The
  subsystems were centered at~$x=200$. In both (a) and (b), 
  the slopes of the lines corresponding to the same~$f$ 
  values are the same.}
  \label{fig:KLD-dim-extensive}
 \end{figure}

\begin{figure}   % second figure
\caption{
  KLD correlation length~$\xi_{\rm KLD}(f)$ versus
  fraction~$f$ for the data of
  \figref{fig:KLD-dim-extensive}, showing a nonlinear
  dependence over an order of magnitude in $\xi_{\rm
  KLD}(f)$. } \label{fig:xi-KLD-vs-f}
 \end{figure}

\begin{figure}   % third figure
\caption{
  (a) Comparison of the KLD and Lyapunov
  dimension correlation lengths (squares and circles
  respectively) for the non-equilibrium transition of
  the 2d Miller-Huse model, for which the two-point
  correlation length~$\xi_2$ diverges at~$g=0.205$
  \protect\cite{Miller93}.  The values for~$\xi_\delta$
  were taken from Ref.~\protect\cite{OHern96} while the
  values for~$\xi_{\rm KLD}(f)$ were calculated for the
  fraction~$f=0.95$ using spatiotemporal data of the
  Miller-Huse model on a 2D~square lattice of
  size~$L=30$ with periodic boundary conditions.
% Initial data were chosen
% to be uniformly distributed random numbers in the
% interval $[-0.1,0.1]$.
  After a transient time of $2\times 10^6$ iterations,
  $T=10^4$ time-samples of~$S=L^2$ lattice sites in the
  range~$12^2 \le S \le 20^2$ were used to define the
  data matrix~$\bf A$.  The double-headed arrow
  indicates the range in parameter~$g$ used to study
  weakly inhomogeneous dynamics in
  \figref{fig:xi-KLD-spatial-variation}. }
  \label{fig:xi-kld-parametric}
\end{figure}

\begin{figure}   % fourth figure
\caption{
  (a) KLD correlation length~$\xi_{\rm KLD}(f)$
  with~$f=0.95$ (solid circles) versus spatial
  coordinate~$i$ for a weakly inhomogeneous 2D
  Miller-Huse model on a square lattice with periodic
  boundary conditions, with rectangular geometry
  $L_x=300$ and~$L_y=30$. The solid circles were
  obtained by studying the extensive scaling of~$D_{\rm
  KLD}(f)$ for subsystems of size ~$L_x=9$, $11$, $13$,
  and~$15$ centered on the particular coordinate~$i$.
  The dashed curve in (a) represents the value
  of~$\xi_{\rm KLD}(f)$ interpolated from
  \figref{fig:xi-kld-parametric}, corresponding to an
  infinite nontransient homogeneous system with
  constant value~$g(i)$ corresponding to that
  particular location in space.  The scatter matrix was
  calculated from $T=10^4$ time-samples after
  waiting~$2\times 10^6$ time units for transients to
  decay. (b) The coupling constant~$g(i)$ varied
  spatially in the~$x$-direction according to the
  equation $g=0.17 + 0.02 \sin(2\pi i/300)$. }
  \label{fig:xi-KLD-spatial-variation}
 \end{figure}

\centerline{\epsfysize=8.5in \epsfbox{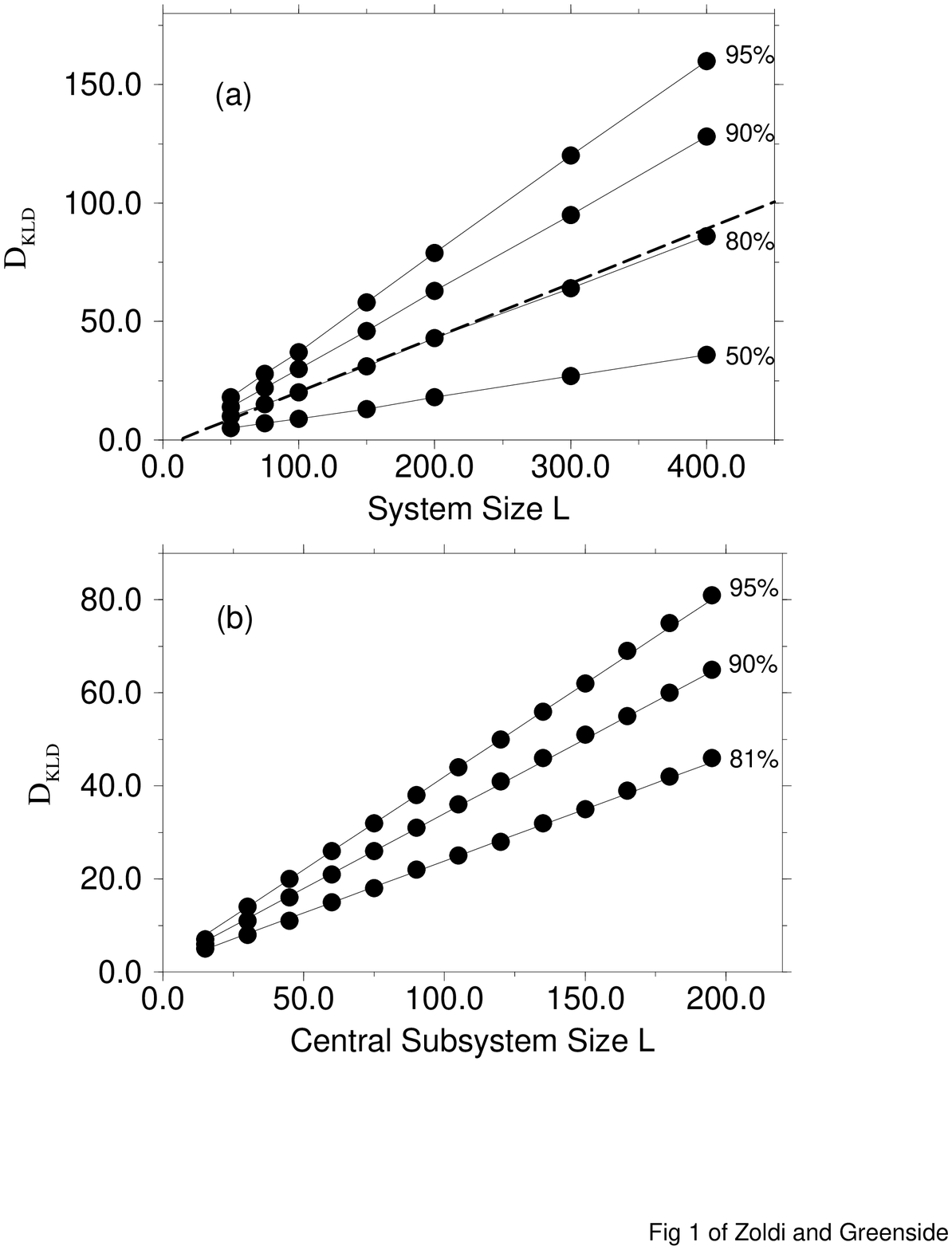}}
\centerline{\epsfysize=8.5in \epsfbox{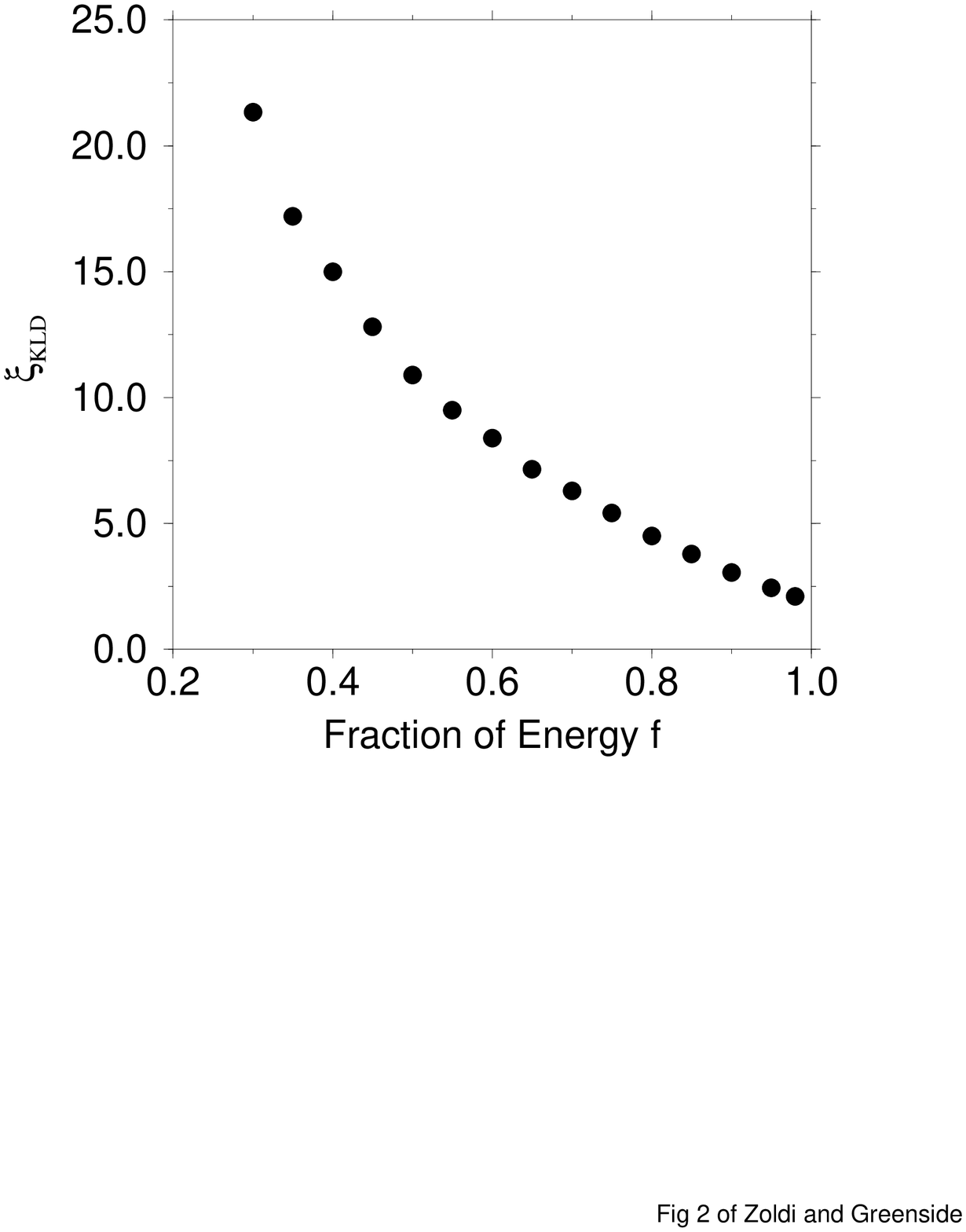}}
\centerline{\epsfysize=8.5in \epsfbox{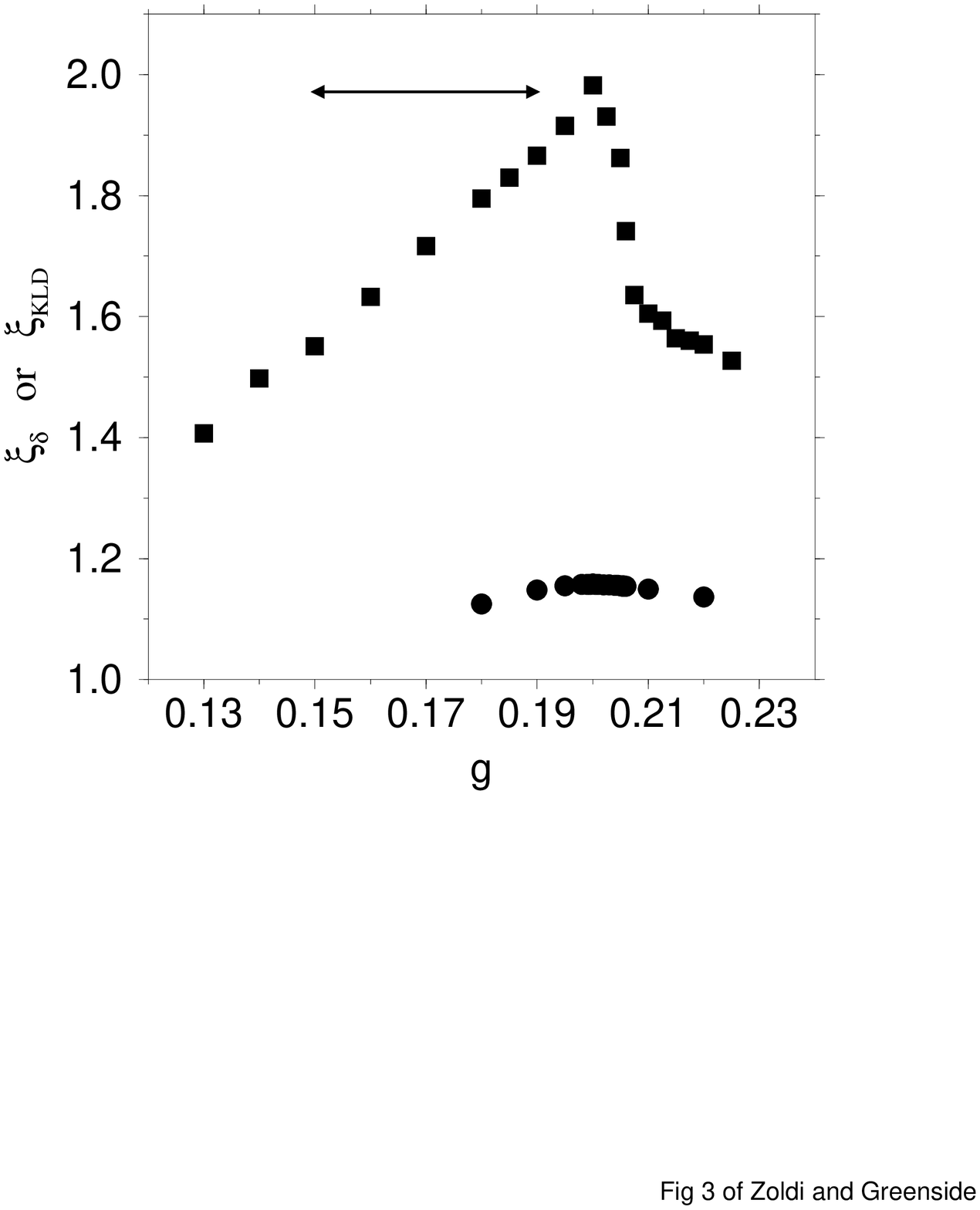}}
\centerline{\epsfysize=8.5in \epsfbox{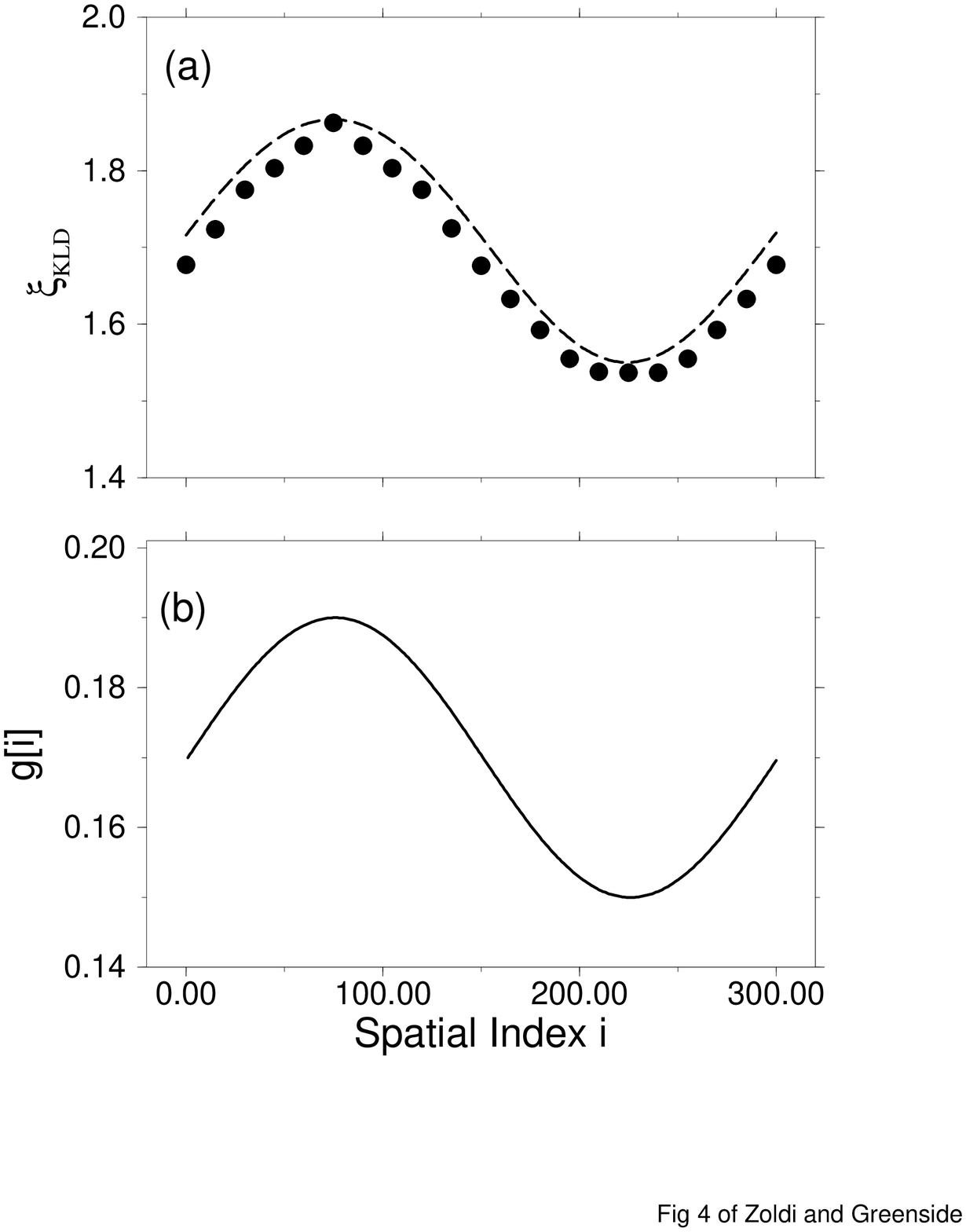}}

\end{document}